\documentclass[manuscript,screen]{acmart}
\AtBeginDocument{%
  \providecommand\BibTeX{{%
    \normalfont B\kern-0.5em{\scshape i\kern-0.25em b}\kern-0.8em\TeX}}}

\setcopyright{none}
\copyrightyear{2023}
\acmYear{2023}
\acmDOI{}

\acmConference[CHI '23]{Conference on Human Factors in Computing Systems}{April 23--28,
  2023}{Hamburg, Germany}
\acmBooktitle{CHI 2023: Workshop on Privacy Interventions and Education (PIE): Encouraging Privacy Protective Behavioral Change Online,
 April 28, 2023, Hamburg, Germany} 
\acmPrice{}
\acmISBN{978-1-4503-XXXX-X/18/06}

\begin{document}

\title{Revisiting the Design Agenda for Privacy Notices and Security Warnings}

\author{Diana Korka}
 \orcid{0000-0003-2019-1599}
 \email{diana.korka@unil.ch}
 \affiliation{%
  \institution{University of Lausanne}
  \country{Switzerland}}

\author{Kavous Salehzadeh Niksirat}
 \orcid{0000-0003-4438-3544}
 \email{kavous.salehzadehniksirat@unil.ch}
\affiliation{%
 \institution{University of Lausanne}
  \country{Switzerland}}

 
\author{Mauro Cherubini}
 \orcid{0000-0002-1860-6110}
 \email{mauro.cherubini@unil.ch}
\affiliation{%
 \institution{University of Lausanne}
  \country{Switzerland}}

\renewcommand{\shortauthors}{Korka et al.}

\definecolor{Revisioncolor}{HTML}{0000FF}
\newcommand{\revchi}[1]{\textcolor{black}{#1}}

\begin{abstract}
  System-generated user-facing notices, dialogs, and warnings in privacy and security interventions present the opportunity to support users in making informed decisions about identified risks.
  However, too often, they are bypassed, ignored, and mindlessly clicked through, mainly in connection to the well-studied effect of user fatigue and habituation. 
  \revchi{The contribution of} this position paper \revchi{is to provide a summarized review of} established and emergent design dimensions and principles to limit such risk-prone behavior and \revchi{to} identify three \revchi{emergent research and design} directions for privacy-enhancing dialogs. 
\end{abstract}




\received[revised]{16 April 2023}

\maketitle

\section{Motivation}\label{sec:motivation}

System-generated user-facing notices and dialogs in security and privacy interventions are ubiquitous elements of human-computer interaction, in a multitude of contexts such as websites, mobile applications, computers, smartphones, wearables, privacy assistants or smart home devices. 
From a design perspective, these interventions draw on recommendations of risk communication theory and applicable legal requirements. 
Typically they inform or warn users of existing privacy and security risks and present them with the choice of opting out of that risk, or continuing should they have sufficient evidence or should they choose to ignore warnings. 
They are, for example, privacy notices informing on practices with users' personal data and requesting users' consent, dissuasive privacy-related dialogs identifying possible privacy breaches and offering alternatives for opting out, software installation dialogs, warnings for ``your connection is not private'', or warn-and-continue dialogs for downloading/saving email attachments \revchi{(see Figures~\ref{fig:mozaic_privacy} and ~\ref{fig:mozaic_security} in Appendix~\ref{SuppB})}. 
They differ from interventions that design out the identified risk (e.g., no personal data is collected at all) or those that completely prevent users from taking the risky option (e.g., Internet service providers blocking the use of specific IP addresses).

While such mechanisms for privacy and security interventions can be beneficial by educating users to protect themselves against potential risks, numerous experimental field and laboratory studies show that users routinely ignore them, thus exposing themselves to the negative consequences of identified risks ~\cite{dhamijaWhyPhishingWorks2006, ebertBolderBetterRaising2021, obarBiggestLieInternet2018, schaubDesigningEffectivePrivacy2017, schechterEmperorNewSecurity2007, wuSecurityToolbarsActually2006, bohmeTrainedAcceptField2010}. 
The observed behavior brings into question the effectiveness of such interventions in everyday life.  
We reviewed the related literature to identify design recommendations and unexplored research areas for supporting user behavior. Particularly, we looked at three goals for rendering the design more effective: 
(i)~improve the informative and educational value of such notices to help users understand the risk they expose themselves \revchi{and/or others} to, 
(ii)~strengthen their persuasion power such that users take the safe option when justified, and 
(iii)~limit habituation and inattentive click-through behavior. 
Specifically, we sought to identify common guidelines to inform the design of notices and dialogs,\footnote{Notices and dialogs can vary greatly from one device to another. For example, for wearables, privacy notices are often made available upon first use, and on an accompanying website, as displaying such information on a small screen is not feasible.} because small differences in design have been shown to impact their effectiveness significantly~\cite{akhaweAliceWarninglandLargescale2013}.

\section{Background}\label{sec:background}

\textbf{Privacy policies or notices} are intended to help users make informed data privacy decisions. 
Nevertheless, consumers often find themselves clueless when presented with notices in the form of long lists of data practices as they are written in highly technical and legalistic terms, vague to remain flexible, and aiming primarily to demonstrate compliance to regulators~\cite{bohmeTrainedAcceptField2010, schaubDesigningEffectivePrivacy2017}. 
\citet{obarBiggestLieInternet2018} found that 74\% of the respondents to a large-scale survey clicked through and skipped reading privacy policies, 97\% agreed to them, and  many failed to recognize when some of these notices included problematic and harmful data practices. 
\citet{ebertBolderBetterRaising2021} assessed the effectiveness of short and concise privacy notices from a fitness tracking app by asking participants to recall data practices and confirmed that users rarely read privacy notices when these are hidden behind links.

\textbf{Security warnings} identify a risk to those exposed to it, afford users the choice to avoid it, and explain the consequences of risk materialization~\cite{wogalterPurposesScopeWarnings2006}. 
Enhancing the effectiveness of security warnings is a well-researched topic, yet risk-prone user behavior persists. 
\citet{wuSecurityToolbarsActually2006} examined the usability of security toolbars in web browsers and showed them to be ineffective in preventing phishing attacks. 
Similarly, for SSL (or Secure Socket Layer) warnings, more than half of the participants in a lab study ignored them, thus potentially exposing themselves to fraudulent websites~\cite{dhamijaWhyPhishingWorks2006}. 
In another study, 53\% of participants ignored security warnings on an online banking site, clicked through, and entered passwords, thus potentially exposing themselves to man-in-the-middle or phishing attacks~\cite{schechterEmperorNewSecurity2007}. 

The potential benefits of enhancing the effectiveness of such mechanisms are three-fold: 
(i)~to improve their informative and educational function, 
(ii)~to enhance their persuasive power to orient users towards safer options, and 
(iii)~to address \emph{habituation}, also known as \emph{``the repetition suppression effect, for repeated exposure''}~\cite{andersonUsersArenNecessarily2014}. 
\citet{andersonUsersArenNecessarily2014} used functional magnetic resonance imaging (fMRI) to measure habituation in participants exposed to security warning messages. 
They showed that repeated exposure leads to significantly reduced neural activity in the brain's visual processing center, and attention diminishes.  
The unconscious automatic response tunes out learned repetitive stimuli to focus attention on another task at hand. 
Response decay was already noticeable from the second presentation and continued to increase linearly with the number of repetitions; at the same time, the habituation response was found to be significantly greater for security warnings than for general software applications. 
%
%
Habituation can have a contamination effect across computer systems. 
If users become accustomed to ignoring security warnings in one context, they may apply the same behavior in another similar context. 
For example, \citet{sunshineCryingWolfEmpirical2009} observed that participants remembered previous security warnings and applied them to other websites, similar to \citet{egelmanYouVeBeen2008} who showed that participants who recognized a warning message from a trusted website were significantly less likely to read a similar message when it appeared on another website and thus also less likely to heed it. 

All is not lost however, as attention recovers spontaneously after some time of non-exposure to the habituation-provoking stimulus~\cite{amranHabituationEffectsComputer2018}. 
In addition, attention decay can be countered by \emph{dishabituation} interventions, such as exposure to polymorphic warnings, with repeated changes in the message appearance~\cite{andersonUsersArenNecessarily2014}. 


\section{The Design Space for Privacy Notices and Security Warnings}\label{sec:design_space}

Prior work identified a set of \revchi{design dimensions}, guidelines and principles, by considering privacy notices and security warnings in isolation. 
We summarize the design dimensions for \textbf{privacy notices} in Figure~\ref{fig:summary} in Appendix~\ref{SuppA}, the left hand side. 
\citet{schaubDesignSpaceEffective2018} identify four dimensions for the design space by reviewing the usability literature. 
\begin{itemize}  
    \item \textbf{Timing.} The moment the notice is triggered in the design flow, for example, at setup, periodically, persistent, or on demand.
    \item \textbf{Channel.} The means of providing the notice, for example, within the same interaction platform of the system, or rather through a secondary channel, for example, through video tutorials, or yet through a public channel, such a privacy beacon, which broadcasts preferences to any nearby users.
    \item \textbf{Modality.} The form the notice takes, for example visual, auditory, haptic, machine-readable.
    \item \textbf{Control.} The level of control afforded by the notice to the user, whether the notice is in a take-it-or-leave-it form, whether it enables users to refine their privacy preferences, and whether consent is required before accessing the service.
\end{itemize}


Furthermore, the authors discuss recommendations for privacy notice design, in connection to the four dimensions. 
For timing, two considerations stand out: 
(i)~seeking better integration of the notice into a system's interaction flow and 
(ii)~considering that reading the notice conflicts with the user's primary task. 
For channel, the preference is, when possible, the direct (main) channel (confirmed by ~\citet{ebertBolderBetterRaising2021}), providing the notice in the same platform, application or device the user interacts with, thus avoiding context change, to say a separate accompanying website with privacy terms and conditions. 
For modality, a combination of, for example, text and image content can help attract attention and better convey the message. In addition, further consideration is given to reducing the level of reading difficulty of the text, avoiding the use of specialized terms or jargon, and preferring a simple and concise layout of notices over lengthy texts.
For control, the recommendation goes towards affording choice, similar to the pre-selected checkboxes for the data types to be shared designed by \citet{wangOnlineExperimentPrivacy2013}.

\revchi{\citet{fengDesignSpacePrivacy2021} extend the design space for affording choice, where choice is defined as communicating and providing access to actions that result in an effective control over data practices. For example, they refine the Control dimension of \citet{schaubDesignSpaceEffective2018}, to \textit{Type} and \textit{Functionality of choices}. Type considers the comprehensiveness of choices: binary choice (e.g., opt-in or opt-out), multiple choice (e.g., several opt-ins, for more granular data practices), contextualized choice (e.g., before a particular risk occurs), and more comprehensive rights-based choice (e.g., affording data portability, and rectification). Functionality considers the way the system presents privacy choices, the enforcement mechanism, and the feedback on privacy choice status.} 

Another consideration is that privacy notices tend to serve both the purpose of demonstrating compliance with legal requirements\footnote{For example, General Data Protection Regulation (GDPR)~\cite{theeuropeanparliamentandthecounciloftheeuropeanunionRegulationEU20162016} in Europe or California Consumer Privacy Act (CCPA)~\cite{officeofthecaliforniaattorneygeneralCaliforniaConsumerPrivacy2018} in the US.}, and of effectively informing users of current data practices. 
Thus, \citet{schaubDesigningEffectivePrivacy2017} argue in favor of a new practice, complementing long, complex, compliance-oriented privacy notices, with user-centric just-in-time, tailored, multi-layered, and specific notices that are better integrated into the interaction flow. 
We did not retain a specific recommendation on design to reduce habituation to privacy notices. 

For \textbf{security messages}, \citet{wogalterPurposesScopeWarnings2006} provided design principles for several aspects of a warning, including identifying standard components of a warning message, standard text color related to the level of risk, background, and symbols to be used in a signal word panel (with reference to ANSI~\cite{nationalelectricalmanufacturersassociationANSIZ535320222022}), message text format, font type, wording, pictorial symbols, giving the warning with adequate time to enable the user to act, recommending usability testing, and suggesting to change the appearance of the warning to reduce habituation (i.e., polymorphic design). 
Furthermore, the usable security team at Microsoft proposed a list of guiding principles for the design of effective security warnings~\cite{reederPosterHelpingEngineers2011} (summarized in Figure~\ref{fig:summary} in Appendix~\ref{SuppA}, the right-hand-side). 
\begin{itemize}
    \item \textbf{Necessary.} A warning should only interrupt the interaction flow when a user absolutely needs to be involved.
    \item \textbf{Explained.} It should explain the decision the user needs to make and give them helpful information for making that choice. Therefore, the warning should identify the source application that triggered it, the process the user needs to undergo to make the right decision, the risk consequences, the unique knowledge that the user has to make the decision, the list of available choices, and the information evidence to be used in the decision-making.
    \item \textbf{Actionable.} It should only be presented if the user can realistically make a better-informed choice.
    \item \textbf{Tested.} It should undergo usability testing.
\end{itemize}

Several usability tests shed further insights on the design space. 
Few privacy studies centered on how to make the information and educational content easier to grasp. 
For example, \citet{kelleyStandardizingPrivacyNotices2010, reinhardtVisualInteractivePrivacy2021, kitkowskaOnlineTermsConditions2022} leverage the advantages of a \emph{compact and structured layout} to study the impact of different privacy policy formats, with exciting prospects for nutrition-label tabular summaries, and for adding \textit{visual icons}, or other graphical representations to convey \textit{comparable information}. \citet{windlAutomatingContextualPrivacy2022} took on these recommendations to develop a tool that automatically extracts and displays concise and comparable contextual privacy notices from existing privacy notices, leading to better-informed behavior.
Most interventions focused on how to attract user attention to the information. 
Affording more \textit{granular control} over the data types shared, through separate checkboxes, resulted in lower privacy permissions~\cite{wangOnlineExperimentPrivacy2013}, and incorporating \textit{interaction elements} (e.g., swiping) 
improved attention~\cite{karegarDilemmaUserEngagement2020}. 
Others successfully explored \textit{nudges} for privacy, in online social media, by drawing attention to the audience of a future post or by delaying the post briefly and allowing additional edits~\cite{wangFieldTrialPrivacy2014} and by adding examples of the personal data surrendered when accepting permissions for a mobile application~\cite{harbachUsingPersonalExamples2014}. 
For security, proven nudges consisted of interactive attractors to the information evidence to be used in decision-making~\cite{bravo-lilloYourAttentionPlease2013}, claiming the presence of a human auditor in the loop to audit security decisions~\cite{brustoloniImprovingSecurityDecisions2007}, or persuading users to take the safer decisions using more visually appealing design and through dissimulating unsafe options behind ``advanced'' buttons, requiring additional interaction steps~\cite{feltImprovingSSLWarnings2015}. 
A few security interventions addressed habituation, by varying the form in which users provide inputs~\cite{brustoloniImprovingSecurityDecisions2007}, or by animating warning messages (e.g., jiggle) and varying the size of the warning icons~\cite{andersonHowPolymorphicWarnings2015}. 

\section{An Emerging Research and Design Agenda}\label{sec:agenda}

We identify future research and design \revchi{directions} for security and privacy user-facing dialogs. 
These can apply to the growing number of initiatives to improve the effectiveness of such interactions and reduce the extent to which users ignore, bypass, or become habituated to them. We propose these ideas for discussion at the workshop on Privacy Interventions and Education (PIE): Encouraging Privacy Protective Behavioral Change Online, of the ACM CHI Conference on Human Factors in Computing Systems (2023)\footnote{Workshop on Privacy Interventions and Education (PIE): Encouraging Privacy Protective Behavioral Change Online, ACM CHI Conference on Human Factors in Computing Systems, 23-28 April 2023, Hamburg, Germany \href{https://programs.sigchi.org/chi/2023/program/session/102064}{https://programs.sigchi.org/chi/2023/program/session/102064}}.

\subsection{Leveraging Behavioral Theories and Dual-System Thinking}

Most existing privacy- and security- dialogs support rational decision-making. 
We ask, \textit{how can behavioral decision-making theories and dual-system thinking be leveraged to underpin research on improved effectiveness?}

While most previous efforts to enhance the effectiveness of interventions focused on supporting users in rational decision-making, dual system thinking theory~\cite{kahnemanThinkingFastSlow2012} shows that human decision-making \revchi{is only boundedly rational, suffers from cognitive biases}, and can be underpinned by emotions, values, cognitive frames, and worldviews. 
\revchi{\citet{redmilesDancingPigsExternalities2018} measured that rationality accounted for between 48 to 56\% of the decisions in a behavioral economics-based experiment, in which participants were required to make security decisions with clearly identified benefits and costs. Arguably with privacy notices, benefits and costs may not be immediately apparent.}

A number of studies have begun to explore the design space in this area. 
For example, \citet{andersonUsersArenNecessarily2014} found that security warnings trigger more fear and anxiety signals in the brain as compared to general software applications. 
But it is not clear how these emotions mediate attention and habituation. 
\citet{cherubiniWhenForcingCollaboration2021} designed fear-, shame-, and empathy-arousing dissuasive warnings in a context of multi-party privacy risks, informed by dual-system theory and with the intention of provoking self-reflection. 
Further research would be needed to understand the impact of such warnings in the wild. 
Furthermore, some studies explored the potential of comic-based design complements in attracting users' attention to the privacy policy content, with encouraging results~\cite{tabassumIncreasingUserAttention2018, kitkowskaEnhancingPrivacyVisual2020}. \citet{kitkowskaEnhancingPrivacyVisual2020} used the Antecedents -- Privacy Concerns -- Outcomes (APCO) decision-making model and dual-process theory to inform the design of  privacy notices and found that the affective state (positive or negative) elicited by a privacy note, can have an indirect effect on the decision to disclose information. 
Curiosity was also associated with improved privacy comprehension. In addition, providing users with control over privacy decisions leads to higher satisfaction when combined with curiosity-enhancing cues. 
These few studies show the potential of further exploring \textit{emotion-arousing design} in attracting attention.

Analyzing simplified lines of reasoning for observing cues and arriving at conclusions can be a valuable analytical tool. 
Identifying mental models can help inform more effective designs that address common misconceptions or knowledge gaps. 
For example, \citet{bravo-lilloBridgingGapComputer2011} found that novices use principally a website's appearance to judge its trustworthiness. 
Another methodological approach, participatory design can help understand users' perspectives, and involve them in the design process (see, for example,~\citet{salehzadehniksiratThoughtYouWere2021}). For example, after repeated exposure, participants reported becoming habituated to the prompts of a multi-party privacy mediation bot, but at the same time, familiarity helped them feel more confident in interacting with it \cite{salehzadehniksiratPotentialMediationChatbots2023}.

\subsection{Further Exploring the Design Space of Sustainable and Persuasive Dialogs}

We ask, \textit{what design elements can sustainably persuade users to make more informed risk decisions?} 
\textit{Under what conditions can users take on the habit of making safe choices by default? }

Given users' limited technical understanding, it is difficult to inform them thoroughly of the risk threat at hand. 
Perhaps behavior change theories can be leveraged to guide users towards less risky choices. 
Several studies have already started exploring the potential of nudges~\cite{acquistiNudgesPrivacySecurity2017} to enhance privacy (e.g.,~\cite{choeNudgingPeopleAway2013, wangFieldTrialPrivacy2014}) and security interactions (e.g.,~\cite{bravo-lilloYourAttentionPlease2013, feltImprovingSSLWarnings2015}), with mixed results for user satisfaction. 
Equally important to comprehend are dark patterns (i.e., nudging users towards riskier choices)~\revchi{~\cite{nouwensDarkPatternsGDPR2020}}. 
Mapping out more comprehensively the design space of nudging interventions would help, as they currently only target two aspects of the design space: 
(i) increase the salience of information evidence to be considered by users before risk decisions, 
(ii) frame risk-related decision alternatives as main (safe) and secondary (unsafe).

Additionally, it is interesting to understand how standard privacy and security dialogs compare in terms of compliance with self-determination theory (SDT)~\cite{ryanSelfDeterminationTheoryFacilitation2000} by exploring aspects related to users' autonomy (i.e., feeling in control), competence (i.e., feeling competent to decide), and relatedness (i.e., feeling connected to others), to evaluate to what extent these interventions are capable of instilling a lasting behavioral impact on users. 
Some studies (e.g.,~\cite{ertanCyberSecurityBehaviour2020}) have already addressed the impact of social/peer influence in changing security-related behavior. 
Going further, one could investigate, for example, how social influence can be leveraged to enhance privacy notices and dialogs. 
And given the influence of advice, under what conditions, if any, can users take on the habit of by default selecting less risky actions (e.g., choosing not to consent to share private data with websites).

More research could shed light on how to use design to instill behavior change in privacy interventions. 
For example, when investigating alternatives for privacy policies' layout, \citet{kelleyStandardizingPrivacyNotices2010} found that preserving a standard tabular nutrition-label-inspired format was helpful because it rendered the comparison across policies possible. 
We wonder whether, with repeated exposure, users could take on the habit to read such policies because the comparison is afforded.

\subsection{Integrating the Design Space for Privacy and Security Dialogs}

Finally, inspired by a study from \citet{feltImprovingSSLWarnings2015}, we ask, \textit{can design principles for \revchi{security risk communications be used to inform the design of privacy notices}?} 

\citet{feltImprovingSSLWarnings2015} took into account how their newly designed security warning fitted with the wider service context, and especially with the other warnings generated by the platform (e.g., by considering the relative level of danger of different security risks). 
Section~\ref{sec:design_space} enumerates design features and guidelines that were arrived at by considering privacy notices and security warnings in isolation. We argue that comparing the design space for privacy notices with that for security warnings can help identify potential gaps.
Both design spaces are underpinned by recommendations of risk communication theory and have a good deal of elements in common. \revchi{In addition, whereas for privacy dialogs, only design dimensions and features are identified, for security warnings, (voluntary) design principles are available.}
%
 Some of the design principles outlined for security warnings, may have not yet been fully considered in the design space for privacy notices, and may be relevant when designing privacy notices (e.g., highlighting the risk consequences and the process the user should follow to make less risky choices). Figure~\ref{fig:summary} in Appendix~\ref{SuppA} attempts to compare the two design spaces and to identify common areas of partial overlap for similar design features (identified with dotted rectangles), as well as design features given prominence under security, but which are not fully reflected in the privacy design space (identified with an asterisk and in italics).


\bibliographystyle{ACM-Reference-Format}
\bibliography{habituation}

\appendix
\begin{figure}[ht]
    \section{Appendix}\label{SuppA}
    \bigskip
	\centering
	\includegraphics[width=\textwidth]{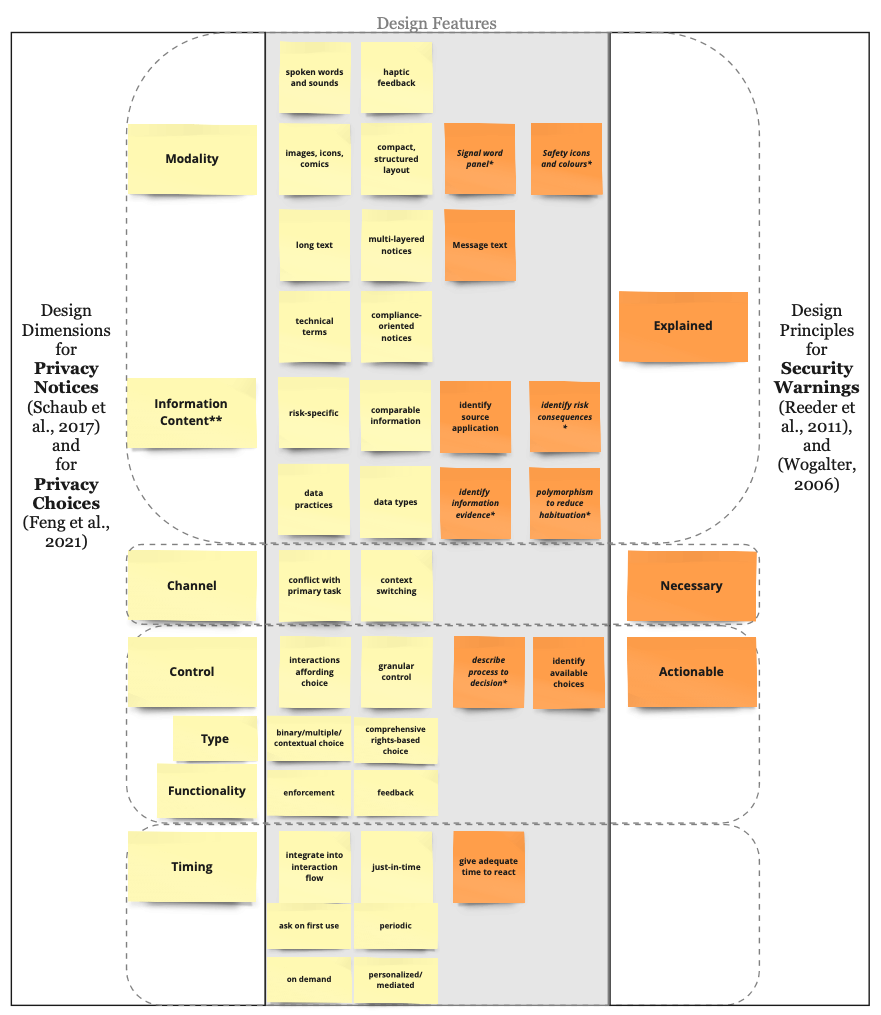}
	\caption{Outline of design dimensions, principles and features for privacy notices and security warnings}
\label{fig:summary}
\footnotesize * refers to features from security warning design not fully reflected in the privacy notice design space and\\ ** refers to a suggested separate dimension of the design space.
\end{figure}
\begin{figure}[ht]
    \section{Appendix}\label{SuppB}
    \bigskip
	\centering
	\includegraphics[width=\textwidth]{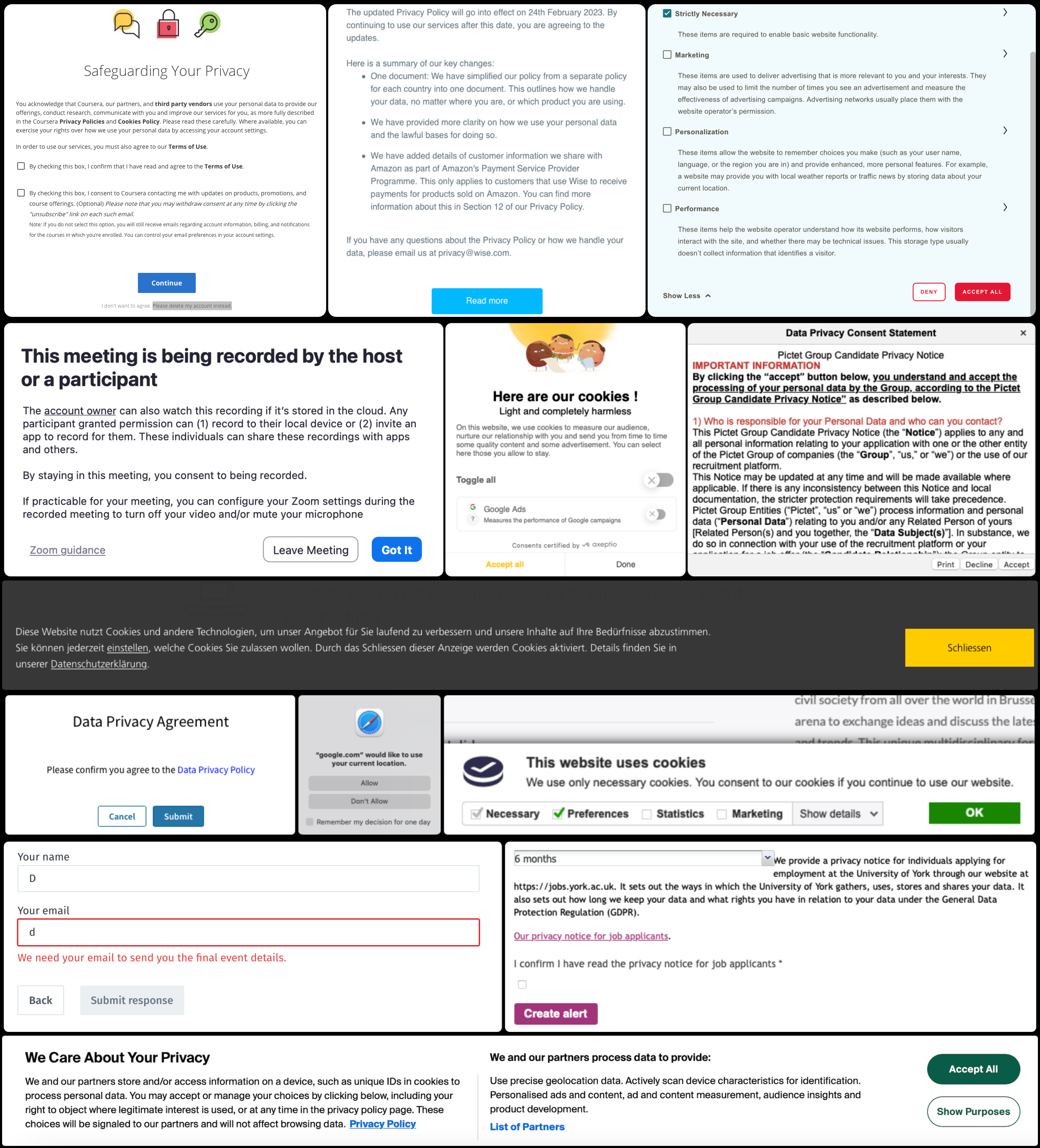}
	\caption{Mosaic of privacy notices examples}
\label{fig:mozaic_privacy}
\end{figure}

\begin{figure}[ht]
    \bigskip
	\centering
	\includegraphics[width=\textwidth]{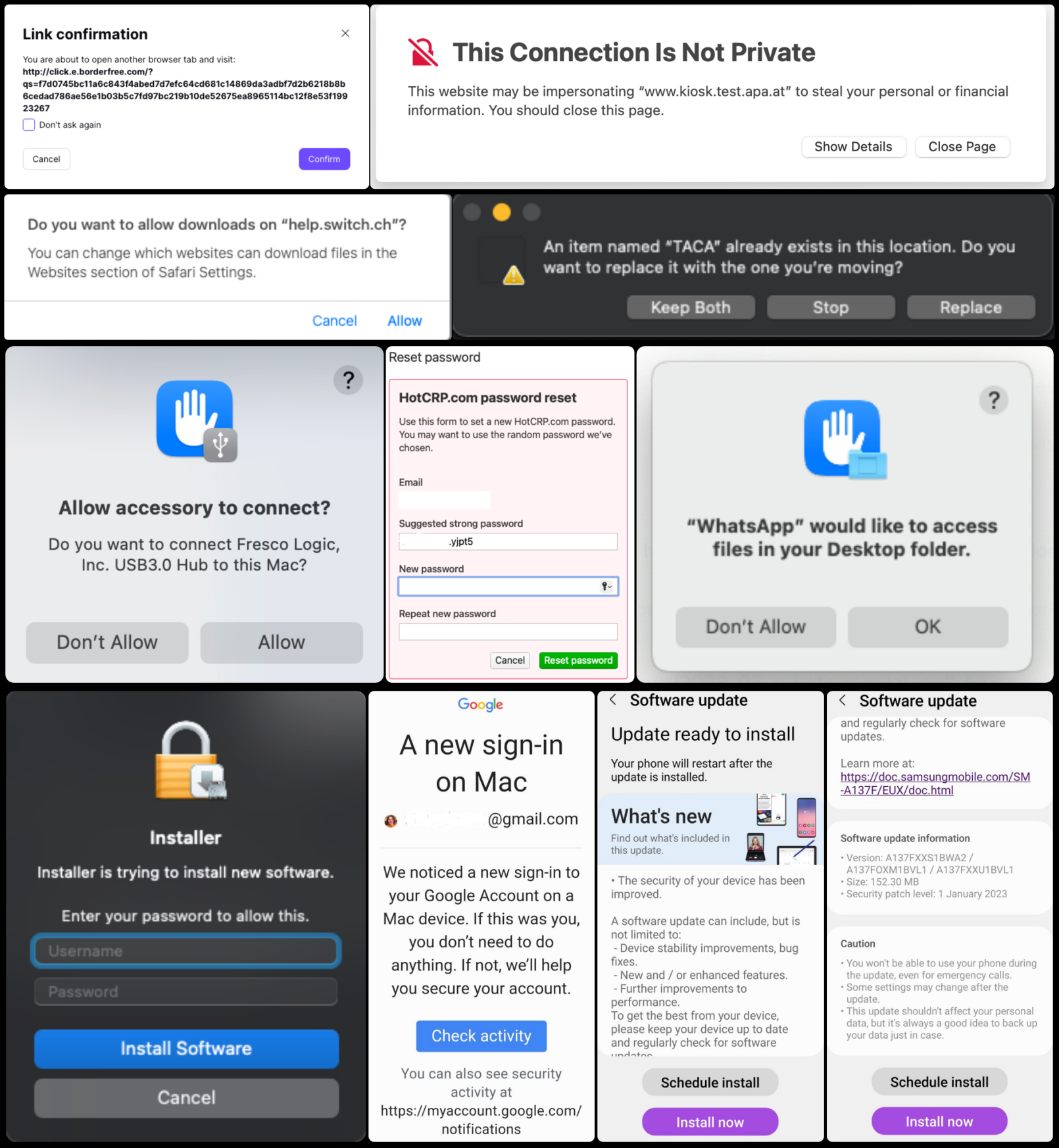}
	\caption{Mosaic of security warnings examples}
\label{fig:mozaic_security}
\end{figure}

\end{document}